\documentclass[twocolumn,showpacs,showkeys,preprintnumbers,amssymb,aps,superscriptaddress,prb]{revtex4}
\usepackage{graphicx}
\usepackage{dcolumn}
\usepackage{bm}
\usepackage{epsfig}

\begin{document}

\preprint{APS/123-QED}

\title{Exact Ground States of Frustrated Spin-1 Ising-Heisenberg and Heisenberg Ladders in a Magnetic Field}

\author{Jozef Stre\v{c}ka}
\affiliation{Institute of Physics, Faculty of Science, P. J. \v{S}af\'{a}rik University, Park Angelinum 9, 040 01, Ko\v{s}ice, Slovakia}
\author{Fr\'ed\'eric Michaud}
\affiliation{Institute of Theoretical Physics, Ecole Polytechnique F\'{e}d\'{e}rale de Lausanne (EPFL), CH-1015 Lausanne, Switzerland}
\author{Fr\'ed\'eric Mila}
\affiliation{Institute of Theoretical Physics, Ecole Polytechnique F\'{e}d\'{e}rale de Lausanne (EPFL), CH-1015 Lausanne, Switzerland}
\date{\today}

\begin{abstract}
Ground states of the frustrated spin-1 Ising-Heisenberg two-leg ladder with Heisenberg intra-rung coupling and only Ising interaction along legs and diagonals are rigorously found by taking advantage of local conservation of the total spin on each rung. The constructed ground-state phase diagram of the frustrated spin-1 Ising-Heisenberg ladder is then compared with the analogous phase diagram of the fully quantum spin-1 Heisenberg two-leg ladder obtained by density matrix renormalization group (DMRG) calculations. It is demonstrated that both investigated spin models exhibit quite similar magnetization scenarios, which involve intermediate plateaux at one-quarter, one-half and three-quarters of the saturation magnetization. 
\end{abstract}



\maketitle

\section{Introduction}
\label{sec:intro}

Over the last few decades, quantum spin ladders have been actively studied mainly in connection with spin-liquid behaviour, quantum critical points and superconductivity under hole doping of some cuprates (see Ref. [\onlinecite{bat07}] for a review). In particular, the frustrated spin-1/2 Heisenberg two-leg ladder exhibits a striking dimerized ground state [\onlinecite{gel91}] and a low-temperature magnetization process with an intermediate plateau and magnetization jumps [\onlinecite{hon00}]. 

Another challenging topic of current research interest consists of the theoretical investigation of related models such as the quantum spin-1 Heisenberg two-leg ladder [\onlinecite{cha06,mic10}]. The main goal of the present work is to find the exact ground states of a simpler spin-1 Ising-Heisenberg ladder and to contrast them with the respective ground states of the pure quantum spin-1 Heisenberg ladder. Note that the former model is analytically tractable using the procedure developed in Refs. [\onlinecite{bar98,bar09,roj13}] and it brings insight into the relevant behaviour of the latter not fully integrable model.

\section{Frustrated Ising-Heisenberg Ladder}
\label{sec:IHL}

Consider first the frustrated spin-1 Ising-Heisenberg ladder with the Heisenberg intra-rung interaction and the unique Ising interaction along the legs and diagonals. 
The total Hamiltonian of the investigated model is given by 
\begin{eqnarray}
\hat{\cal H} = \sum_{i=1}^{N} \left[ J \hat{\textbf{S}}_{1,i} \cdot \hat{\textbf{S}}_{2,i} \right.
                                   \!\!\!&+&\!\!\! J_{1} (\hat{S}_{1,i}^{z} + \hat{S}_{2,i}^{z}) (\hat{S}_{1,i+1}^{z}+\hat{S}_{2,i+1}^{z})  \nonumber \\
                                   \!\!\!&-&\!\!\! \left. h (\hat{S}_{1,i}^{z} + \hat{S}_{2,i}^{z}) \right],
\label{ham}
\end{eqnarray}
where $\hat{\textbf{S}}_{\alpha,i} \equiv (\hat{S}_{\alpha,i}^{x}, \hat{S}_{\alpha,i}^{y}, \hat{S}_{\alpha,i}^{z})$ denotes spatial components of the spin-1 operator, the former suffix $\alpha$ = 1 or 2 enumerates the leg and the latter suffix specifies a lattice position within a given leg. The coupling constant $J$ denotes the isotropic Heisenberg intra-rung interaction, the parameter $J_1$ determines the Ising interaction along the legs and diagonals, $h$ is an external magnetic field. 

For further convenience, let us introduce the spin operator $\hat{\textbf{T}}_{i} = \hat{\textbf{S}}_{1,i} + \hat{\textbf{S}}_{2,i}$, which corresponds to the total spin angular momentum of the $i$th rung. It can be easily proved that the operators  $\hat{\textbf{T}}_{i}^2$ and  $\hat{T}_{i}^z$ commute with the Hamiltonian (\ref{ham}), i.e. $[\hat{\cal H}, \hat{\textbf{T}}_{i}^2] = [\hat{\cal H}, \hat{\textbf{T}}_{i}^z] = 0$, which means that the total spin of a rung and its $z$ component represent conserved quantities with well defined quantum numbers. The complete energy spectrum of the frustrated spin-1 Ising-Heisenberg ladder then readily follows from the relation
\begin{eqnarray}
E = - 2 N J + \frac{J}{2} \sum_{i=1}^{N} T_{i} (T_{i} + 1) + J_{1} \sum_{i=1}^{N} T_{i}^{z} T_{i+1}^{z} - h \sum_{i=1}^{N} T_{i}^{z},
\label{diag}
\end{eqnarray}
which depends just on the quantum numbers $T_i = 0,1,2$ and $T_i^z = -T_i, T_i+1, \ldots, T_i$ determining the eigenvalues of the total spin of the $i$th rung and its $z$th spatial projection, respectively. Using this procedure, the spin-1 Ising-Heisenberg two-leg ladder has been rigorously mapped to some classical chain of composite spins and accordingly, we can readily find all available ground states by looking for the lowest-energy state of Eq. (\ref{diag}).

\section{Frustrated Heisenberg Ladder}
\label{sec:HL}

Next, we will also consider the frustrated spin-1 Heisenberg two-leg ladder defined by the Hamiltonian
\begin{eqnarray}
\hat{\cal H} = \sum_{i=1}^{N} \left[ J \hat{\textbf{S}}_{1,i} \cdot \hat{\textbf{S}}_{2,i} \right.
                          \!\!\!&+&\!\!\! J_{1} (\hat{\textbf{S}}_{1,i} + \hat{\textbf{S}}_{2,i}) \cdot (\hat{\textbf{S}}_{1,i+1} \cdot \hat{\textbf{S}}_{2,i+1}) \nonumber \\
                          \!\!\!&-&\!\!\! \left. h (\hat{S}_{1,i}^{z} + \hat{S}_{2,i}^{z}) \right],
\label{hamh}
\end{eqnarray}
which represents the pure quantum analogue of the frustrated spin-1 Ising-Heisenberg ladder discussed previously. Taking advantage of the definition for the total spin of each rung, the Hamiltonian (\ref{hamh}) of frustrated spin-1 Heisenberg two-leg ladder can be rewritten into the form
\begin{eqnarray}
\hat{\cal H} = - 2 N J + \frac{J}{2} \sum_{i=1}^{N} \hat{\textbf{T}}_{i}^2 
             + J_{1} \sum_{i=1}^{N} \hat{\textbf{T}}_{i} \cdot \hat{\textbf{T}}_{i+1} - h \sum_{i=1}^{N} \hat{T}_{i}^{z}.
\label{diagh}
\end{eqnarray}
According to Eq. (\ref{diagh}), the frustrated spin-1 Heisenberg ladder can be rigorously decomposed into the direct sum of quantum spin chains with spin 0, 1 or 2 at each site. The ground state of such a system can be shown to be either a homogeneous chain, with the same spin at all sites, or a chain with alternating spins on every other site. Comparing the energy of the different chains, obtained either analytically or using DMRG simulations, the exact ground-state phase diagram of the frustrated spin-1 Heisenberg ladder can be constructed [\onlinecite{cha06,mic10}].

\section{Results and Discussion}
\label{sec:res}

\begin{figure}
\begin{center}
\includegraphics[width=0.49\textwidth]{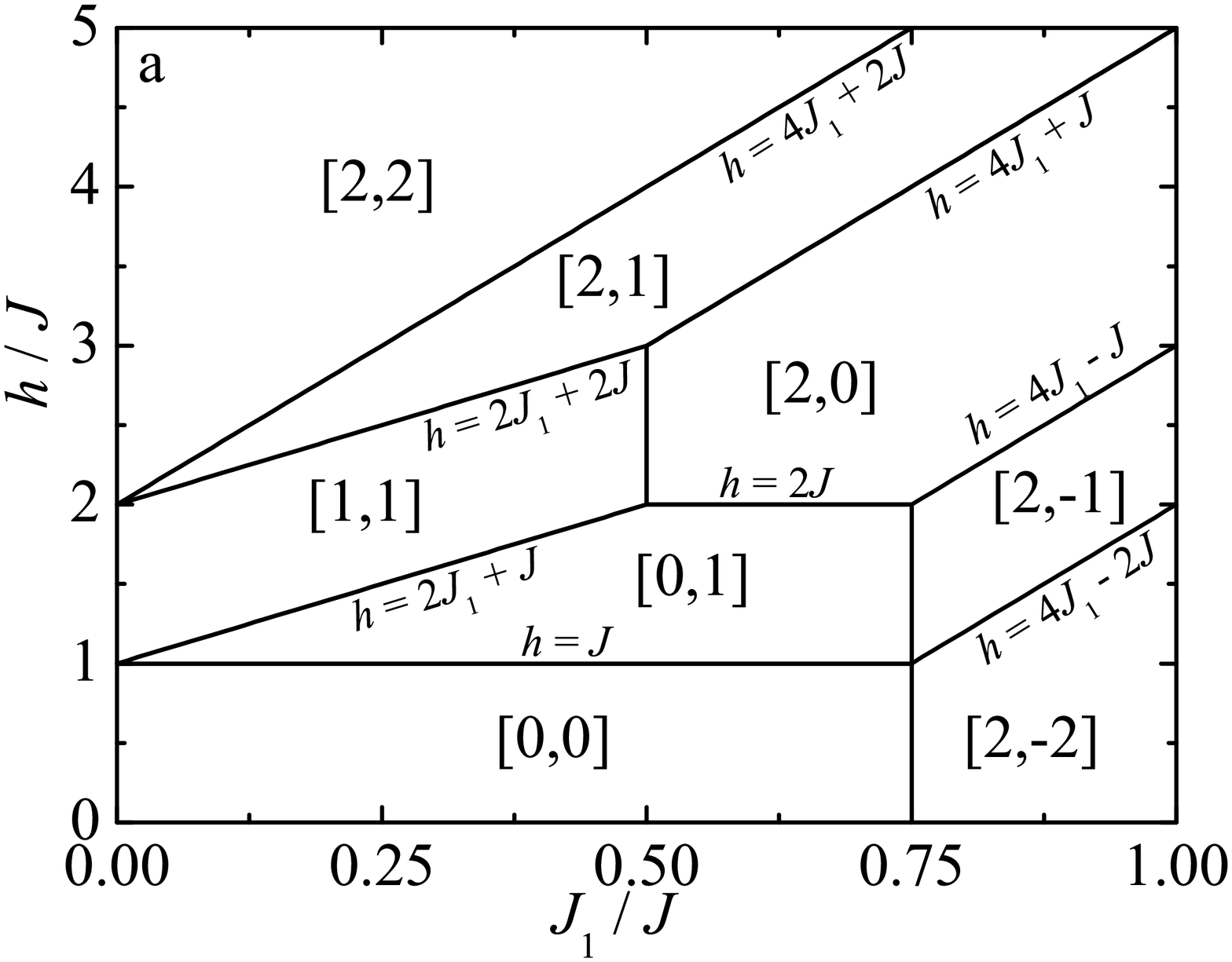}
\includegraphics[width=0.49\textwidth]{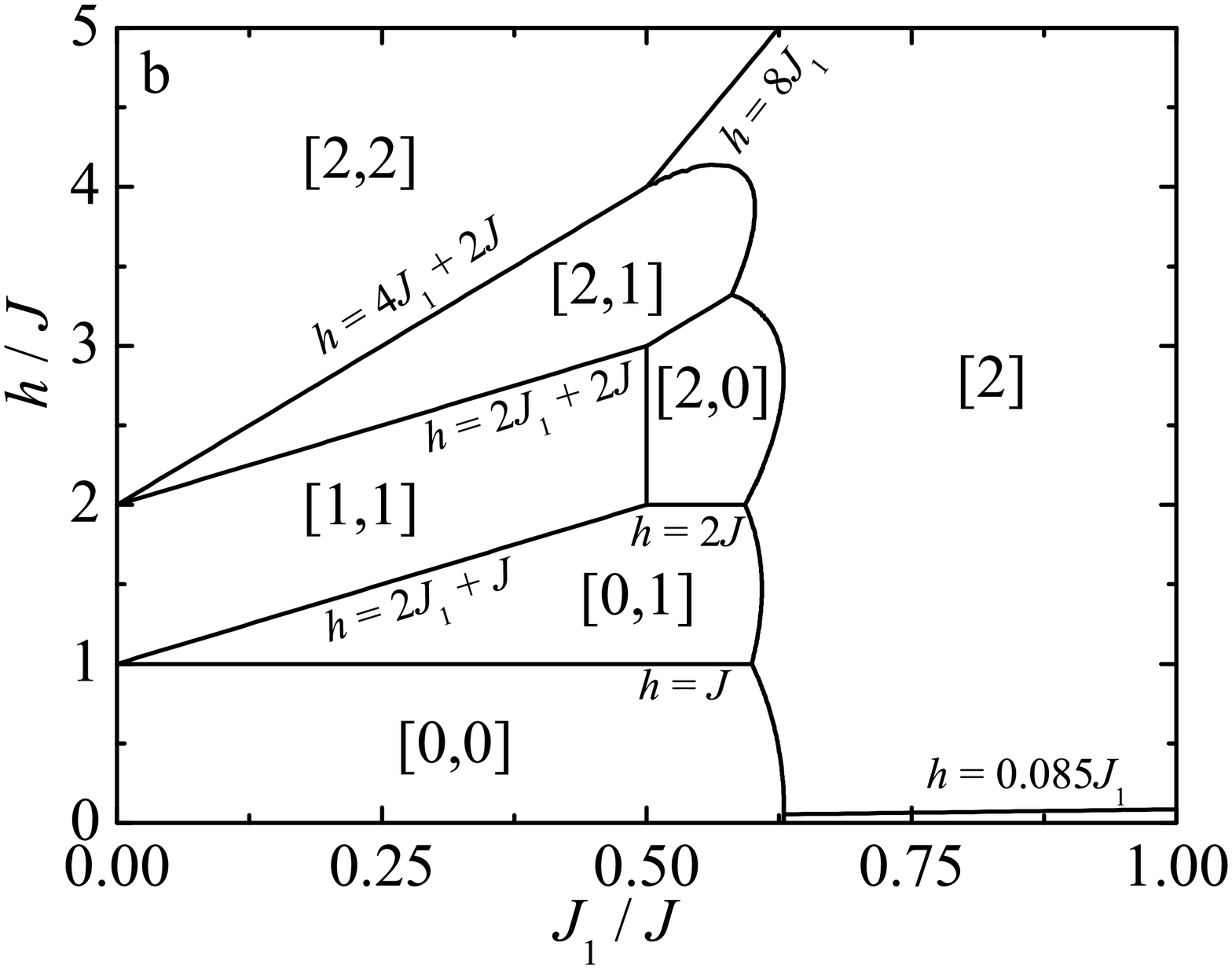}
\end{center}
\vspace{-0.7cm}
\caption{Ground-state phase diagrams of the frustrated spin-1 ladder described within: (a) the Ising-Heisenberg model; (b) the pure Heisenberg model. 
For details see the text.}
\label{fig1}
\end{figure}

The constructed ground-state phase diagrams of the frustrated spin-1 Ising-Heisenberg and Heisenberg ladders are depicted in Fig.\ref{fig1}a and Fig.\ref{fig1}b, respectively. The ground states of the spin-1 Ising-Heisenberg ladder can be discerned according to the $z$th projection of the total spin on two consecutive rungs $[T_i^z,T_{i+1}^z]$, because $T_i = |T_i^z|$ holds for all available ground states. The quantum ground states [0,0], [0,1], [1,1], [2, 1] and [2,0] represent six different phases, whereas 
$T_i = T_i^z = 0$ implies a formation of two singlets on the $i$th rung, $T_i = |T_i^z| = 1$ entails only one singlet and $T_i = |T_i^z| = 2$ denotes fully polarized rungs without singlets. Besides, two classical ground states [2,2] and [2,-2] are pertinent to the ferromagnetic and antiferromagnetic ordering. The magnetization normalized with respect to its saturation equals zero for [0,0] and [2,-2], one-quarter for [0,1] and [2,-1], one-half for [1,1] and [2,0], three-quarters for [2,1] and unity for [2,2]. Altogether, it can be concluded that the frustrated spin-1 Ising-Heisenberg ladder always exhibits a stepwise magnetization curve, which involves intermediate plateaux at one-quarter, one-half and three-quarters of the saturation magnetization that are however of different origin. 

It is quite clear from Fig.\ref{fig1}b that the ground-state phase diagram of the pure quantum Heisenberg ladder exactly coincides with that of the Ising-Heisenberg ladder just for sufficiently weak inter-rung interactions $J_1/J \leq 0.5$. A relatively good agreement between both phase diagrams is still observed in the parameter space $0.63 \geq J_1/J \geq 0.5$, where the gapless phase [2] with a continuously varying magnetization is present between the intermediate plateaux instead of direct magnetization jumps. The gapless phase [2] corresponds to the Luttinger-liquid phase of the effective spin-2 quantum Heisenberg chain. Finally, the gapped Haldane phase of the effective spin-2 quantum Heisenberg chain emerges for $J_1/J \geq 0.63$ at sufficiently low fields.   

In conclusion, we have rigorously found the ground states of the frustrated spin-1 Ising-Heisenberg and Heisenberg ladders in a magnetic field. It has been demonstrated that the Ising-Heisenberg ladder always exhibits a stepwise magnetization curve with three different intermediate plateaus, while the same quantum ground states can be identified in the pure quantum Heisenberg ladder provided the intra-rung coupling is sufficiently large.  

\begin{acknowledgments}
This work was financially supported by the projects VEGA 1/0234/12 and APVV-0132-11. F. Michaud thanks S.R. Manmana for providing his DMRG code.
\end{acknowledgments}

\end{document}